\documentclass[aps,pra,amsmath,amssymb,twocolumn,superscriptaddress,showpacs]{revtex4}  

\usepackage{bbm}
\usepackage{graphicx}
\usepackage{dcolumn}
\usepackage{bm}
\usepackage{amsmath}
\usepackage{amssymb,amsthm}

\usepackage{hyperref}

\usepackage[sans]{dsfont}

\newcommand{\ket}[1]{|#1\rangle}
\newcommand{\braket}[2]{\langle{#1}|{#2}\rangle}

\newcommand{\bra}[1]{\langle#1|}

\def\eq{\begin{eqnarray}}
\def\en{\end{eqnarray}}

\def\beq{\begin{eqnarray}}
\def\een{\end{eqnarray}}
\def\bfig{\begin{figure}}
\def\efig{\end{figure}}

\begin{document}

\title{Double-Fock Superposition Interferometry for Differential Diagnosis of Decoherence}
\author{Malte C. Tichy}
\address{Department of Physics and Astronomy, University of Aarhus, DK--8000 Aarhus, Denmark}
\author{Young-Sik Ra}
\address{Department of Physics, Pohang University of Science and Technology (POSTECH), Pohang, 790-784, Korea}
\author{Hyang-Tag Lim}
\address{Department of Physics, Pohang University of Science and Technology (POSTECH), Pohang, 790-784, Korea}
\author{Clemens Gneiting} 
\address{Physikalisches Institut der Albert-Ludwigs-Universit\"a{}t, D--79104 Freiburg, Germany}
\author{Yoon-Ho Kim}
\address{Department of Physics, Pohang University of Science and Technology (POSTECH), Pohang, 790-784, Korea}
\author{Klaus M\o{}lmer}
\address{Department of Physics and Astronomy, University of Aarhus, DK--8000 Aarhus, Denmark}

\begin{abstract}
 Interferometric signals are degraded by decoherence, which encompasses dephasing, mixing and any distinguishing which-path information. These three paradigmatic processes are fundamentally different, but, for coherent, single-photon and $N00N$-states, they degrade interferometric visibility in the very same way, which impedes the diagnosis of the cause for reduced visibility in a single experiment. We introduce a versatile formalism for many-boson interferometry based on double-sided Feynman diagrams, which we apply to a protocol for differential decoherence diagnosis: Twin-Fock states $|N,N\rangle$ with $N \ge 2$ reveal to which extent decoherence is due to path distinguishability or to mixing, while double-Fock superpositions $|N:M\rangle = \left( |N,M\rangle + |M,N\rangle \right)/\sqrt{2} $ with $N > M >0$ additionally witness the degree of dephasing. Hence, double-Fock superposition interferometry permits the differential diagnosis of decoherence processes in a single experiment, indispensable for the assessment of interferometers. 
\end{abstract}

\date{\today}
\maketitle

\section{Introduction}
\begin{figure}[th] \center
\includegraphics[width=\linewidth,angle=0]{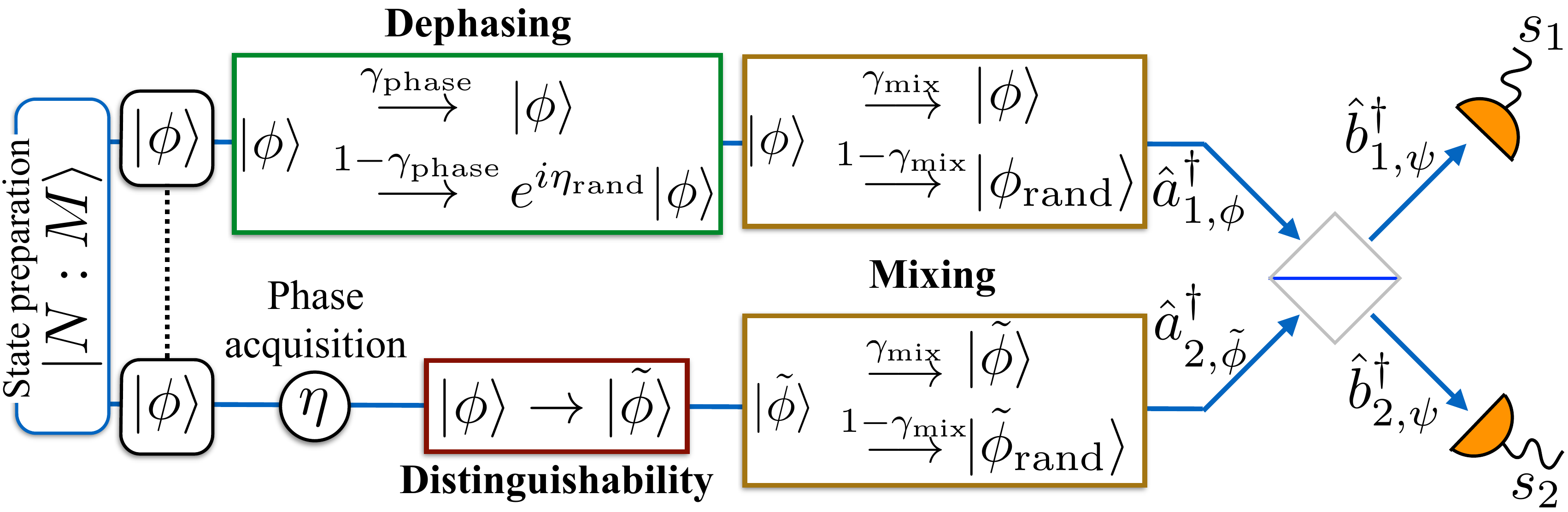}
\caption{Interferometer subject to  decoherence. Particles in either arm start in the same internal single-particle state $\ket{\phi}$. In the lower arm, each particle  acquires the phase $\eta$, which is measured by combining the arms at a beam splitter and recording the number of particles $s_1$ and $s_2$ in the two detectors. The interferometric measurement is jeopardized by dephasing (a random phase $\eta_{\text{rand}}$, acquired here  in the upper arm with probability $1-\gamma_{\text{phase}}$), path distinguishability (a change in the internal states of the particles travelling through the lower arm, resulting in a finite scalar product $|\braket{\phi}{\tilde \phi}|=\gamma_{\text{dist}}$ and leading to decoherence \cite{ratichycomment}) and mixing (with probability $1-\gamma_{\text{mix}}$ for each arm, the particles are left in an unknown randomly chosen state). 
 }  \label{Interferometer.pdf}
 \end{figure}

The coherent superposition of  physically exclusive single- or many-particle path amplitudes 
 is exemplified best with the minimalistic paradigm of a single particle prepared in the state $\ket{1:0}\equiv (\ket{1,0}+ \ket{0,1})/\sqrt{2}$, a coherent superposition of the upper and lower arms of a Mach-Zehnder interferometer. Fringes appear in the output signal, the probability to find the particle in the upper detector, as a function of the relative phase $\eta$ that the particle acquires between the two arms (Fig.~\ref{Interferometer.pdf}). We define the fringe visibility of the output signal $(s_1,s_2)=(1,0)$,
\eq 
\mathcal{V}^{\ket{1:0}}_{(1,0)} = \frac{\mathcal{P}^{\ket{1:0}}_{(1,0),\text{max}} -\mathcal{P}^{\ket{1:0}}_{(1,0),\text{min}} }{\mathcal{P}^{\ket{1:0}}_{(1,0),\text{max}} +\mathcal{P}^{\ket{1:0}}_{(1,0),\text{min}} } , \label{visibility10}
\en
where $\mathcal{P}^{\ket{N:M}}_{(s_1,s_2)}$ is the probability of the signal with $(s_1,s_2)$ detected particles for the bosonic initial state $\ket{N:M}$, defined below. Interference is jeopardized by several deteriorating effects  summarized as  \emph{decoherence} \cite{RevModPhys.75.715,RevModPhys.76.1267,Breuer:2006ud,Buchleitner:2009fk,JoosZeh}, and, in practice, the visibility $\mathcal{V}^{\ket{1:0}}_{(1,0)}$ never reaches unity. 
The interferometer  in Fig.~\ref{Interferometer.pdf} illustrates three prominent mechanisms for decoherence: Dephasing, mixing, and path distinguishability. By tracing out the internal degrees of freedom of the particle and performing the classical average over phase fluctuations, the state of a single particle in the two arms is described by an effective two-state density matrix
\eq 
\hat \varrho = \frac 1 2  \left( \begin{tabular}{cc}
 1 & $e^{i \eta} \mathcal{V}^{\ket{1:0}}_{(1,0)} $\\ 
$e^{-i \eta}\mathcal{V}^{\ket{1:0}}_{(1,0)} $&  1  
\end{tabular}
\right) , \label{descriptionrho}
\en
where the visibility $\mathcal{V}^{\ket{1:0}}_{(1,0)}$ coincides with the classical ensemble-average of the scalar product $\braket{\phi}{\tilde \phi}$ of the states of the particle in the upper and the lower arms, and thereby quantifies the coherence between the arms. The three decoherence effects are, thus, not differentiated in practice by the single-particle interferometric signal. For -- possibly strong -- coherent states, for which the signal intensity replaces the event probability in Eq.~(\ref{visibility10}), there is no possibility for a qualitative differentiation of decoherence mechanisms either, nor for $N00N$-states, as we show below. 

The parsimonious description inherent to (\ref{descriptionrho}) is certainly sufficient to \emph{predict} the combined impact of dephasing, distinguishability and mixing on an interferometer   \cite{Rauch1999,RevModPhys.84.157}, and an impressive level of understanding of decoherence in nature has been achieved through studies that essentially monitor only the visibility, as demonstrated for large molecules \cite{RevModPhys.84.157}.  
 However, the precise cause of decoherence remains unknown in circumstances that are not well understood, and three  interferometers that are affected by three qualitatively different decoherence processes may exhibit the very same signal. For the characterization and eventual alleviation of decoherence, \emph{differential diagnosis}, i.e.~detailed information about the nature of decoherence, is crucial: Only a well-characterized cause of the signal deterioration  can be thoroughly addressed and eventually removed. 

For example, in nuclear magnetic resonance, it is crucial to distinguish truly irrevocable dephasing from inhomogeneous spin precessions, which can be diagnosed and reversed by spin-echo measurements \cite{SpinDynamics}. In the present context of interferometry, the key to a more differentiated picture of decoherence processes lies in the complexity inherent to entangled many-boson states. Here, we introduce a versatile treatment of bosonic double-Fock superpositions, which encompasses single-particle states, $N00N$-states and double-Fock states of the form $\ket{N,M}$ as special cases. The formalism naturally allows us to treat mixed states and to thereby incorporate decoherence processes such as mixing, distinguishability and dephasing. As a result, a four-particle double-Fock state $\ket{2,2}$ allows a clear diagnosis of  mixing against distinguishability, while the  entangled double-Fock superposition $\ket{2:1} \equiv (\ket{1,2}+\ket{2,1})/\sqrt 2$  additionally quantifies dephasing. States with larger particle numbers promise an even more detailed revelation of the processes that deteriorate interference. 

\section{Double-Fock interferometry}
\subsection{Pure states}
We consider double-Fock superpositions of the form 
\eq
\ket{N:M} &=&\frac{1}{\sqrt 2} \left( \ket{N}^{\phantom 0}_{1, \phi } \ket{M}^{\phantom 0}_{2, \tilde \phi} +  \ket{M}^{\phantom 0}_{1, \phi  } \ket{N}_{2, \tilde \phi} \right)  \label{mnnmstate} \\
\nonumber &\equiv& \frac{1}{\sqrt 2 } \left(\ket{N,M} + \ket{M,N} \right) ,
\en
where $0\le M < N$, and $\ket{K}_{l, \theta}$ denotes the Fock state of $K$ bosons in the interferometric arm $l$ and in the internal state $\ket{\theta}.$  The latter pools all remaining relevant degrees of freedom  by which particles can possibly be distinguished (besides the mode number): For a photon,  $\ket{\theta}$ typically describes the polarization and the spatio-temporal mode function.  Double-Fock superpositions  comprise single-photon ($N=1,M=0$)  and $N00N$-states ($N>1, M=0$) as special cases. 

After propagation through the interferometer, the component $\ket{M,N}$ in (\ref{mnnmstate}) acquires the relative phase  ${(N-M)\eta}$ with respect to $\ket{N,M}$. This phase can be inferred by combining the two arms at a beam-splitter and measuring the probability $\mathcal{P}^{\ket{N:M}}_{(s_1,s_2)}$ to find ${(s_1, s_2=N+M-s_1)}$ particles in the two output modes \cite{Laloe:2010uq}.  In practice, the paths might not be fully indistinguishable, which is reflected by a non-unity scalar product $\braket{\phi}{\tilde \phi}$, i.e.~by partial distinguishability, which complicates the computation of event rates. One approach to partially distinguishable bosons consists in replacing the ideal bosonic permanent by a sum of more general immanants \cite{Tan:2013ix,de-Guise:2014yf}. Alternatively, the initial state can be decomposed into orthogonal components of different degrees of distinguishability \cite{TichyFourPhotons,Ra:2013kx,younsikraNatComm,tichyTutorial}. Neither method, however, offers a straightforward extension to mixed states --  a prerequisite in our context -- because the resulting expressions for the event probabilities feature a complicated dependence on the scalar product $\braket{\phi}{\tilde \phi}$.

\begin{figure}
\includegraphics[width=\linewidth,angle=0]{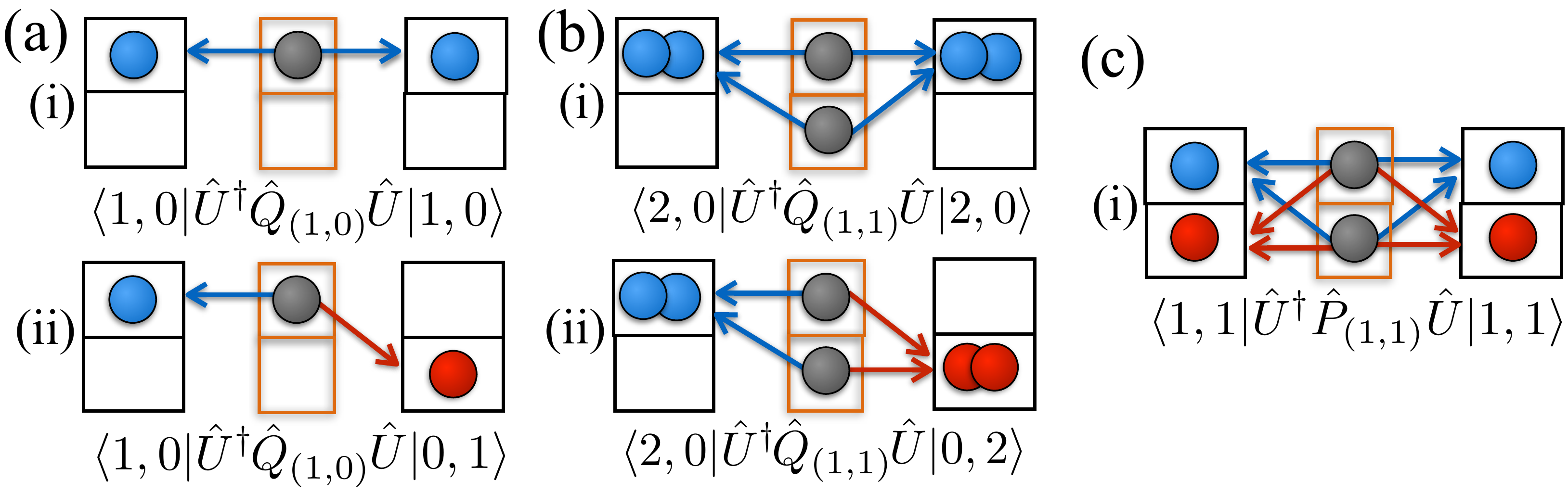}
\caption{  Full double-sided Feynman diagrams, illustrated for (a) $\ket{1:0}$, $(s_1,s_2)=(1,0)$; (b) $\ket{2:0}, (s_1, s_2)=(1,1)$; (c) $ \ket{1,1}$, $(s_1,s_2)=(1,1)$. The initial state is propagated in time by $\hat U$, projected onto $\hat Q_{(1,1)}$ (orange frame, the projector does not differentiate the internal state, hence the gray coloring) and propagated back via $\hat U^\dagger$. The upper rows correspond to the term (i) in Eq.~(\ref{generalexpressionprobability}), the lower rows to (ii); the latter is absent for the twin-Fock state $\ket{1,1}$. For states with possible bosonic exchange processes, there are several competing paths; in general, all paths need to be summed up. }  \label{FirstFeynmanDiagrams.pdf}
\end{figure}

Here, we overcome this shortcoming 
 by treating the coherent many-particle propagation via double-sided Feynman diagrams  \cite{PhysRevA.41.6485}. Our starting point is the expectation value of the projector $\hat Q_{(s_1,s_2)}$, whose eigen-space is defined by the desired particle numbers in the output modes of the beam-splitter [Fig.~\ref{Interferometer.pdf}], 
\eq 
\mathcal{P}^{\ket{N:M}}_{(s_1,s_2)}  &=&  \left| \hat Q_{(s_1,s_2)} \hat U \ket{N:M} \right|^2  \label{prob1st} \\ 
&=&   \bra{N:M} \hat U^\dagger \hat Q_{(s_1,s_2)} \hat Q_{(s_1,s_2)}  \hat U \ket{N:M}   \label{forwardbackward}  \\
&=&  \bra{N,M} \hat U^\dagger \hat Q_{(s_1,s_2)} \hat U \ket{N,M} ~~~~~~~~~~~~~~~~~~~~~\text{(i)}  \nonumber   \\
&+ &
  \Re  \left[ e^{i \eta (N-M) } \bra{N,M} \hat U^\dagger \hat Q_{(s_1,s_2)} \hat U \ket{M,N}  \right],~\text{(ii)} \nonumber
\en
where $\hat U$ describes the many-particle beam-splitter transformation in Fock-space, induced by 
\eq 
\hat a^\dagger_{k,\theta} \rightarrow \frac{1}{\sqrt 2 } \left( i \hat b^\dagger_{k, \theta} +  \hat b^\dagger_{3-k, \theta} \right) ,  \label{timeevolution}
\en 
where $k$ refers to the beam splitter mode (see Fig.~\ref{Interferometer.pdf}) and a phase-shift of $\pi/2$ is acquired upon reflection. In deriving Eq.~(\ref{forwardbackward}), we used that, for a balanced beam-splitter,  the  states $\ket{N,M}$ and $\ket{M,N}$ lead to the same event probability. Eq.~(\ref{forwardbackward}) contains two contributions to the event probability: A main contribution (i) for which the bra- and ket-vectors are the same, and a swapped contribution (ii) with different bra- and ket-vectors. These two terms can be interpreted  as double-sided Feyman-diagrams  that combine propagation forwards and backwards in time, implicit in Eq.~(\ref{forwardbackward}): The state $\ket{N:M}$ is  propagated in time via $\hat U$, projected onto the measurement outcome described by the projector $\hat Q$, and propagated back via $\hat U^\dagger$ [see Fig.~\ref{FirstFeynmanDiagrams.pdf}].

Inserting the transformation (\ref{timeevolution}), we identify the permutations of the particles in the modes that yield the same summands.  All the possibilities for distributing the particles among the modes need to be taken into account; using $ _{\phi}\braket{K}{K}_{\tilde \phi} =\braket{\phi}{\tilde \phi}^K$, we write the signal probability as a polynomial in the scalar product $\braket{\phi}{\tilde \phi}$, 
\begin{widetext}
\eq
\mathcal{P}_{(s_1,s_2)}^{\ket{N:M}} &=& 
 \frac{M! N! }{ s_1! s_2!  2^{M+N}} \sum_{J=0}^{M} \mathcal{C}_{J} \left( \overbrace{ |\braket{\phi}{\tilde \phi}|^{2J} }^{\text{(i)}} + 
 \overbrace{ (-1)^{s_1} |\braket{\phi}{\tilde \phi}|^{2(M-J)} \Re \left[ (i \braket{\phi}{\tilde \phi} e^{i \eta} )^{N-M}  
  \right] }^{\text{(ii)}} \right)
 \label{generalexpressionprobability} ,
\en
 where the explicit form of the combinatorial factor $\mathcal{C}_J$ together with an illustration are given in Appendix \ref{eventprobapp}.

\end{widetext}

\begin{figure}[th] \center
\includegraphics[width=.95\linewidth,angle=0]{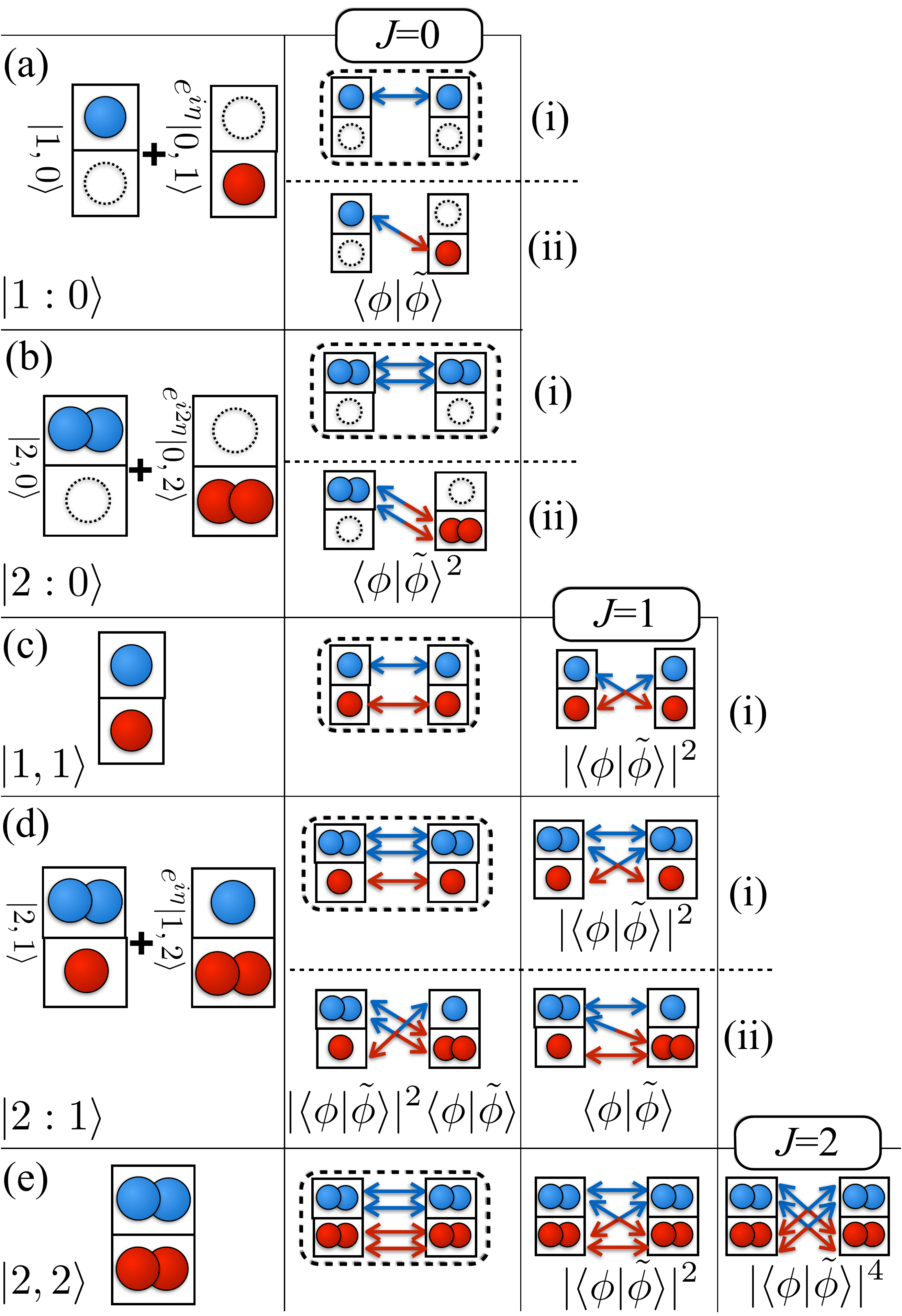}
\caption{ Reduced double-sided Feynman diagrams illustrating  Eq.~(\ref{forwardbackward}), divided up with respect to their contribution to Eq.~(\ref{generalexpressionprobability}): For double-Fock superpositions, the first row corresponds to the phase-independent term in Eq.~(\ref{generalexpressionprobability})(i), the second row denotes the phase-dependent term (ii), naturally absent for twin-Fock-states without phase-dependence. In contrast to Fig.~(\ref{FirstFeynmanDiagrams.pdf}), we omit the intermediate projector and only show the initial state. 
 The columns correspond to processes with different numbers $J$ of bosonic exchange processes. The classical contributions are marked by dotted edges. Particles starting in different modes are possibly distinguishable, reflected by their different colors. Arrows connecting particles of different colors contribute the scalar product $\braket{\phi}{\tilde \phi}$ or $\braket{\tilde \phi}{\phi}$ to the total amplitude.  
 }  \label{Feynmandiagrams.pdf}
 \end{figure}

Eq.~(\ref{generalexpressionprobability}) reveals the dependence of the probability $\mathcal{P}_{(s_1,s_2)}^{\ket{M:N}}$ for the event ($s_1,s_2$) on powers of the indistinguishability parametrized by $\braket{\phi}{\tilde \phi}$ up to order $N+M$.  Absolute-square powers of the form $|\braket{\phi}{\tilde \phi} |^{2 J}$ [term (i)] contribute without any dependence on $\eta$. A particle ``starting'' and ``ending'' in the same arm [horizontal single-colored arrows in Fig.~\ref{Feynmandiagrams.pdf}] contributes a certain amplitude, a particle ending in a different arm [diagonal two-colored arrows] yields an amended amplitude that is attenuated by a factor $\braket{ \phi }{\tilde \phi}$ or $\braket{\tilde \phi}{\phi}$. The many-particle paths for which all particles end in the arm they started from (marked by dotted edges) constitute the ``classical'' contribution, which can be understood via interference-free classical combinatorics. The other terms contain non-vanishing powers of $\braket{\tilde \phi}{\phi}$ and describe different exchange processes. Bosonic exchange processes are those with $J\neq 0$, i.e.~$J$ counts how many particles were actually exchanged between the arms within one component $\ket{N,M}$, such that, naturally, $J \le M$. Exchanges between the $\ket{N,M}$ and $\ket{M,N}$-components lead to a phase-dependence in $\eta$ [term (ii)]. 

For single-photon and $N00N$-states, $M=0$, and the sum (\ref{generalexpressionprobability}) reduces to one term, 
\eq 
\mathcal{P}^{\ket{N:0}}_{(s_1, s_2)} = \mathcal{P}^{\text{dist}}_{(s_1,s_2)} \left[ 1 + 
(-1)^{s_1}\Re \left[ \left( i \braket{\phi}{\tilde \phi} e^{i \eta} \right)^{N}   
  \right]  \right] , \label{N00Nstatesignal}
\en 
where $ \mathcal{P}^{\text{dist}}_{(s_1,s_2)}={N \choose s_1}/2^{N}$ is the ``classical'' combinatorially obtained probability to find $(s_1,s_2)$ distinguishable particles in the output modes. Consistently with $J \le M=0$, no bosonic exchange processes take place.  

For twin-Fock states $\ket{N,N}$, the event probability can also be obtained using Eq.~(\ref{generalexpressionprobability}), but neglecting the second summand (ii), i.e.~the phase-dependent contribution:  Twin-Fock states do not carry any phase-relation between the two modes, therefore, a phase acquired in one arm  manifests itself only as a global, non-observable phase. As a consequence, twin-Fock states are immune to dephasing. The input state $\ket{1,1}$ leads to Hong-Ou-Mandel interference \cite{Hong:1987mz}; for higher occupation, we obtain terms proportional to $|\braket{\phi}{\tilde \phi}|^{2J}$ with $J=0, 1, \dots , N$  \cite{younsikraNatComm}.

Double-Fock superpositions with $0<M < N$ combine the best of both worlds: phase-sensitivity and bosonic effects. Since Eq.~(\ref{mnnmstate}) is a superposition of two two-mode Fock states, bosonic bunching governs the general statistics of the particles in the output modes \cite{Laloe:2010uq,Tichy:2012NJP,tichyTutorial}. Simultaneously, interference between the two components $\ket{N,M}$ and $\ket{M,N}$  permits to measure the phase $\eta$.  The phase-sensitivity of the $\ket{N:M}$-state is enhanced by a factor $N-M$ with respect to the single-photon case, just like for  $N00N$-states \cite{PhysRevLett.85.2733}, which  feature an enhancement of a factor $N$. The richness of interference effects is reflected by the four different contributions depicted in Fig.~\ref{Feynmandiagrams.pdf}(d). 
In general, the state $\ket{N:M}$ leads to $M+N+1$ distinguishable events: $(N+M,0), (N+M-1,1), \dots (0,N+M)$, each of which exhibits a certain dependence on higher powers of the scalar product $\braket{\phi}{\tilde \phi}$.

\subsection{Mixed states}
In the previous section, we took into account the possible deterioration of interference due to path distinguishability ($|\braket{\phi}{\tilde \phi}|\neq 1$), but we assumed that the state of the particles is always the same when they reach the beam splitter. Due to non-unitary random processes, however, we need to assume that the particles in the upper (lower) arm are in the internal state $\ket{\psi_j}$ ($\ket{\tilde \psi_k}$) with probability $p_j$ ($\tilde p_k$), i.e.~in a mixed state. 
One then experiences event probabilities corresponding to the \emph{classical} average (weighted by the $p_j$ and $\tilde p_k$) of the \emph{quantum-mechanical} probability evaluated for $\ket{\psi_j}$ and $\ket{\tilde  \psi_k}$,  
\eq 
\mathcal{P}^{\ket{N:M}}_{(s_1,s_2), {\text{mix}} } = \sum_{ j, k} p_j \tilde p_k \mathcal{P}^{\ket{N:M}_{\psi_j, \tilde \psi_k} }_{(s_1,s_2)} , 
\en
where we made the dependence on $\ket{\psi_j }, \ket{\tilde \psi_k}$ explicit. 

In other words, we can still use Eq.~(\ref{generalexpressionprobability}), but each power of a scalar product 
$ |\braket{\phi }{\tilde \phi}|^{2m} \braket{\phi }{\tilde \phi}^k$ in (\ref{generalexpressionprobability}) needs to be replaced by the \emph{ensemble-averaged scalar product power} (ASPP), denoted by curly brackets $\{ \}$, 
\eq 
\left\{ |\braket{\phi }{\tilde \phi}|^{2m} \braket{\phi }{\tilde \phi}^k   \right \} = \sum_{j,l } p_{j} \tilde p_{l} |\braket{ \psi_{j} }{ \tilde \psi_{l} }|^{2m} \braket{ \psi_{j} }{ \tilde \psi_{l} }^k ,
\en
which mathematically corresponds to a higher moment of the scalar product. 
ASPPs of higher order, ${(m,k) \neq (0,1)}$, play a crucial role in our subsequent analysis: Deteriorating processes affect the ASPPs of different orders in a different way, such that these quantities can encode information about the decoherence process. 

In general, a possibly complex ASPP cannot be written as a function of the single-particle density matrices $\rho$ and $\tilde \rho$ that describe the particles in the upper and lower arm, because the coherences between the arms single out particular bases $\{ \ket{\psi_k} \}, \{ \ket{\tilde \psi_k} \}$. Consider, for example, a qubit-like particle prepared in $\ket{\phi}=\ket{1}$ and a random process that acts on the upper arm, which leaves the qubit in $\ket{0}$ or $\ket{1}$ with probability 1/2. The average scalar product with an unaffected qubit in the lower arm is then $(\braket{0}{1}+\braket{1}{1})/2=1/2$. A process that leaves the qubit in the upper arm in $\ket{+}=(\ket{0}+\ket{1})/\sqrt{2}$ or $\ket{-}=(\ket{0}-\ket{1})/\sqrt{2}$, however, leads to an average scalar product of $(\braket{+}{1}+\braket{-}{1})/2=0$, and, consequently, to a different interference pattern. That is, even though the respective \emph{single}-particle density matrix is the fully mixed state $(\ket{0}\bra{0}+\ket{1}\bra{1})/2 =(\ket{+}\bra{+}+\ket{-}\bra{-})/2= \mathbbm{\hat 1}/2$ in both cases, the average scalar product differs. 

For absolute-squared scalar products of the form $|\braket{\phi}{\tilde \phi} |^{2m}$, the coherences between the arms are irrelevant [Fig.~\ref{Feynmandiagrams.pdf}], which makes the corresponding ASPPs  independent of the single-particle bases, e.g.~for the ensemble-averaged absolute-square of the scalar product, $ \left\{  |\braket{\phi}{\tilde \phi} |^2 \right\} =  \text{Tr}( \rho \tilde \rho )$ .  
The ASPPs of different orders are widely independent of each other; averaged absolute values merely fulfil  
\eq
\{ |\braket{\phi}{\tilde \phi}|^{m} \}^{k/m} \le \{ | \braket{\phi}{\tilde \phi}|^{k} \} \le \{ |\braket{\phi}{\tilde \phi}|^{m} \}   \label{jensen} ,
\en
for every ${k \ge m \ge 1}$, where the lower bound is due to Jensen's inequality and the upper bound follows from $|\braket{\psi_k}{\tilde \psi_j}| \le 1$. 
For two mixed states with the same eigenvectors, $\ket{\psi_k} = \ket{\tilde \psi_k}$ ($[ \rho, \tilde \rho]=0$), 
  the upper bound of (\ref{jensen}) becomes exact; two pure, possibly distinguishable states $\rho=\ket{\phi}\bra{\phi}, \tilde \rho=\ket{\tilde \phi}\bra{\tilde  \phi}$ saturate the lower bound.

\section{Decoherence model}  \label{decomodel}
In general,  non-unitary maps \cite{Breuer:2006ud} that induce decoherence processes in high dimensions can be arbitrarily complicated, reflecting the possibly complex dynamics in the two interferometric arms. Here, we focus on the  decoherence model illustrated in Fig.~\ref{Interferometer.pdf}, which allows us to model the immediate impact of distinguishability, mixing and dephasing via three survival probabilities $\gamma_{\text{dist}}, \gamma_{\text{mix}}, \gamma_{\text{phase}}$, respectively. 

\subsection{Path distinguishability}
Distinguishability has various causes: On the one hand, we consider an observer with a meter initially prepared in the state $\ket{0}_{\text{meter}}$, coupled to the lower arm. 
If the particle takes the upper path, ${\ket{\phi} \ket{0}_{\text{meter}} \rightarrow \ket{\phi} \ket{0}_{\text{meter}}}$; if it takes the lower path, ${\ket{\phi} \ket{0}_{\text{meter}} \rightarrow \ket{\phi} \ket{\beta}}_{\text{meter}}$. Formally, the leakage of which-path information can be accommodated in an amended internal state $\ket{\tilde \phi}$ of the particle in the lower arm that incorporates the meter \cite{bertet2001}, such that $\braket{\phi}{\tilde \phi}:=\braket{0}{\beta}$. On the other hand, mis-alignment of the setup or any other influence on the interferometric arms that permits to distinguish a particle in the upper arm  from a particle in the lower arm via its internal state ($\ket{\phi}$ and $\ket{\tilde \phi}$, respectively) leads to the same effect. 
We neglect here the systematic acquisition of a relative phase between the two arms, which induces a shift of the overall signal in $\eta$, and assume $\braket{\phi}{\tilde \phi}= \gamma_{\text{dist}} \ge 0$. The overall impact of path distinguishability then leads to 
\eq \{ |\braket{\phi}{\tilde \phi}|^{2m} \braket{\phi}{\tilde \phi}^k   \} =\gamma_{\text{dist}}^{2m+k} .  \label{distinguishabilityaffectd} \en

\subsection{Mixing} \label{subsecmixing}
Mixing can be due to classical noise that disturbs the internal state of the particle in an incoherent manner. 
 Here, we model mixing as follows: With probability $\gamma_{\text{mix}}$, all particles in an arm remain unaffected; with probability $(1-\gamma_{\text{mix}})$, all particles are left in an unknown state that is chosen randomly for each run.  
 That is, our mixing process corresponds to the addition of white noise with strength $1-\gamma_{\text{mix}}$, which leads to the following attenuation of the ASPPs: 
\eq 
\{ |\braket{\phi}{\tilde \phi}|^{2m} \braket{\phi}{\tilde \phi}^k \} &\rightarrow& \{ |\braket{\phi}{\tilde \phi}|^{2m} \braket{\phi}{\tilde \phi}^k \} \gamma_{\text{mix}}^2 , \\
\{ |\braket{\phi}{\tilde \phi}|^{2m} \} &\rightarrow& \{ |\braket{\phi}{\tilde \phi}|^{2m} \} \gamma_{\text{mix}}^2 + \frac{1-\gamma_{\text{mix}}^2}{d} , \nonumber 
\en
where $k \ge 1$. The last term proportional to $1/d$ reflects that the average \emph{absolute-squared} scalar product of two random states is finite in a finite-dimensional Hilbert-space, whereas the average \emph{complex} scalar product vanishes due to isotropy; in the following, we assume  $d\rightarrow \infty$, which allows us to neglect the corresponding terms proportional to $1/d$. 
In contrast to distinguishability, which affects different powers in a different manner [Eq.~(\ref{distinguishabilityaffectd})], mixing amends each ASPP of any power by the same factor $\gamma_{\text{mix}}^2$. 

\subsection{Dephasing} 
Dephasing is ubiquitous: To name two examples, unstable optical setups lead to phase fluctuations in photonic experiments, while atomic interferometers are affected by background gas collisions. We incorporate the loss of phase coherence between the arms of the interferometer by assuming that, with probability $\gamma_{\text{phase}}$, all phases remain unaffected; with probability $1-\gamma_{\text{phase}}$, all particles in the lower mode acquire a uniformly random phase $0 \le \eta_{\text{rand}} \le  2 \pi$. The survival rate $\gamma_{\text{phase}}$ is, thus, independent of the number of particles \footnote{This survival mechanism is fully equivalent to the acquisition of a random phase with a finite amplitude at every run of the experiment, but it possesses conceptual and practical advantages for the computation of probabilities.}. Although each value of $\eta_{\text{rand}}$ induces an interference pattern in $\eta$ with high visibility, the origin of that pattern is shifted by $(N-M)\eta_{\text{rand}}$. Since the shift is unknown and varies from run to run, the experimentally observed interference pattern is washed out. 

We account for dephasing by amending the phase-dependent ASPPs, 
\eq 
 \{ |\braket{\phi}{\tilde \phi} |^{2m}  \braket{\phi}{\tilde \phi}^k \} &\rightarrow &   \{ |\braket{\phi}{\tilde \phi} |^{2m}  \braket{\phi}{\tilde \phi}^k \}  \gamma_{\text{phase}}  ,  \nonumber \\
  \{ |\braket{\phi}{\tilde \phi} |^{2m}  \} &\rightarrow &  \{ |\braket{\phi}{\tilde \phi} |^{2m} \} ,  \label{dephasingmodel}  \en
where $k\ge 1$, i.e.~phase-independent terms remain naturally unaffected by dephasing. 

\subsection{Overall impact of decoherence}
The three decoherence mechanisms commute, and the resulting ASPPs after all processes become the product of the survival rates, 
\eq 
 \{ |\braket{\phi}{\tilde \phi} |^{2m} \braket{\phi}{\tilde \phi}^k \} &= &  \gamma_{\text{phase}}  \gamma_{\text{mix}}^2 \gamma_{\text{dist}}^{2m+k}  , \label{impactsp} \\
  \{ |\braket{\phi}{\tilde \phi} |^{2m}  \} &= &  \gamma_{\text{mix}}^2 \gamma_{\text{dist}}^{2m} \label{impactsp2}  ,
\en
for $k \ge 1, m \ge 0$. Summarizing, on the one hand, observable signals $\mathcal{P}^{\ket{N:M}}_{(s_1,s_2)}$ depend on various ASPPs [Eq.~(\ref{generalexpressionprobability})]. On the other hand, different ASPPs reflect the decoherence parameters in a different way [Eqs.~(\ref{impactsp}),(\ref{impactsp2})]. This nourishes the hope that we can observe differences between decoherence mechanisms in many-particle interference signals. 

\section{Decoherence diagnosis}

\subsection{Fringe visibility for single-photon, $N00N$ and twin-Fock states} \label{snf}
\begin{figure}[th] \center
\includegraphics[width=\linewidth,angle=0]{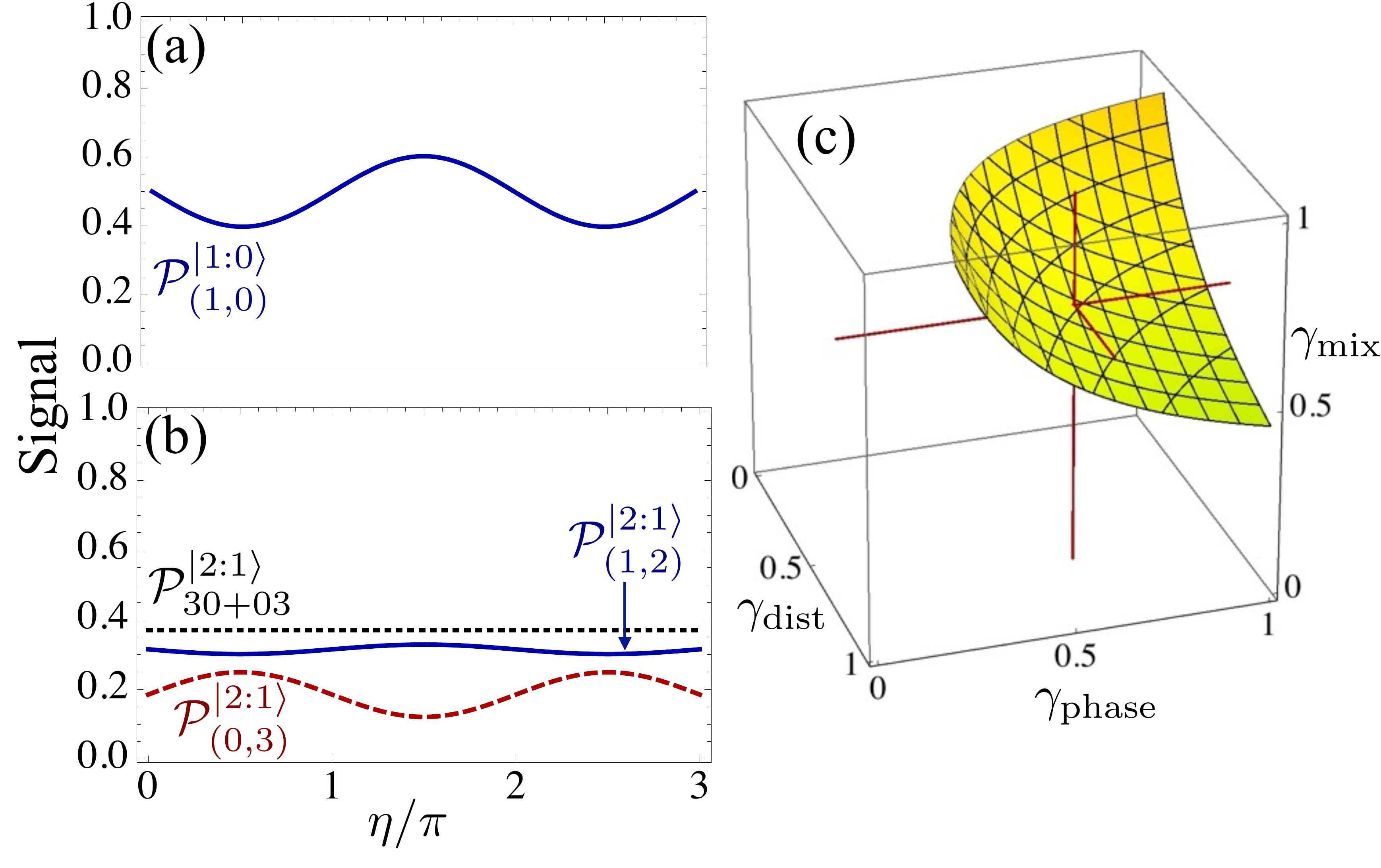}
\caption{(a) Single-particle interference signal with visibility $\mathcal{V}^{\ket{1:0}}_{(0,1)} \approx 0.21$.  (b) In double-Fock superposition interferometry with $\ket{2:1}$, three independent parameters are accessible experimentally. (c) The observed single-particle visibility $\mathcal{V}^{\ket{1:0}}_{(0,1)}$ constrains the three parameters $(\gamma_{\text{dist}}, \gamma_{\text{phase}}, \gamma_{\text{mix}})$ to  a surface defined by Eq.~(\ref{visibility10a}). 
 For the double-Fock superposition $\ket{2:1}$, the three parameters unambiguously determine $\gamma_{\text{phase}}=0.6,\gamma_{\text{dist}}=0.7,\gamma_{\text{mix}}=0.7$ at the crossing point of the three red lines. }  \label{Diagnosis.pdf}
 \end{figure}

The three decoherence mechanisms described in the previous section reduce the fringe visibility of every interferometric signal. For the single-particle state $\ket{1:0}$, the visibility (\ref{visibility10}) is reduced to 
\eq 
\mathcal{V}^{\ket{1:0}}_{(1,0)} &=& 
 \left\{ \braket{\phi}{\tilde \phi}  \right\}  = \gamma_\text{phase}^{\phantom 1}    \gamma_{\text{mix}}^2 \gamma_{\text{dist}}^{\phantom 1}   
,  \label{visibility10a}
\en
i.e.~one can only infer the \emph{product} of definite powers of the three model parameters. Geometrically speaking, a given value of the visibility inferred from the observed signal [Fig.~\ref{Diagnosis.pdf}(a)] leaves room for a surface in the three-dimensional space $( \gamma_\text{phase} , \gamma_{\text{mix}},  \gamma_{\text{dist}} ) $ [Fig.~\ref{Diagnosis.pdf}(c)].

For $N00N$-states $\ket{N:0}$,  only the process in which all particles are exchanged between the modes is relevant [Fig.~\ref{Feynmandiagrams.pdf}(b)], and we find 
\eq 
\mathcal{V}^{\ket{N:0}}_{(s_1,s_2)} =   \left\{ \braket{\phi}{\tilde \phi}^N  \right\}  = \gamma_\text{phase} \gamma_{\text{mix}}^2 \gamma_{\text{dist}}^N   
. \label{visibilityN00N}
\en
 Even though the phase-sensitivity of $\ket{N:0}$ is enhanced with respect to $\ket{1:0}$, there are no contributions from bosonic exchange processes ($J>0$) in Eq.~(\ref{generalexpressionprobability}), and only the \emph{product} of definite powers of the three model parameters can be inferred from the experimental data. 

For the double-Fock state $\ket{1,1}$, the depth of the resulting phase-independent Hong-Ou-Mandel dip \cite{Hong:1987mz} is proportional to 
\eq 
\{ |\braket{\phi}{\tilde \phi}|^2 \}  = \text{Tr}(\rho \tilde \rho) = \gamma_{\text{mix}}^2 \gamma_{\text{dist}}^2,  \label{visiHOM}
\en
which does not allow any differentiation between mixing and distinguishability, which makes the actual purity of a single photon only accessible through advanced analyses \cite{1367-2630-12-11-113052}. In general, two identical mixed states $\rho=\tilde \rho \neq \ket{\phi} \bra{\phi}$ cannot be distinguished from two pure distinguishable states $\ket{\phi}, \ket{\tilde \phi}$ that lead to the same ASPP.

In principle, the visibilities for the three initial states $\ket{1:0}, \ket{2:0}$ and $\ket{1,1}$ depend on the three decoherence model parameters in a different way [compare Eqs.~(\ref{visibility10a}),(\ref{visibilityN00N}) and (\ref{visiHOM})] and, by combining the data from three experiments with different initial states, the  parameters $\gamma_{\text{dist}}, \gamma_{\text{mix}}, \gamma_{\text{phase}}$ can be extracted. However, when the initial state is changed, it is difficult to assess which deteriorating effects are due to the possibly imperfect state preparation and which are caused by the actual decoherence in the interferometer. In the following, we show how to circumvent this problem by extracting the decoherence parameters in one single experiment.

\subsection{Twin-Fock state $\ket{2,2}$}
Twin-Fock states \cite{PhysRevLett.71.1355} with $M=N>1$ clearly bring out the qualitative difference between mixing and distinguishability,  and are routinely generated in the experiment \cite{Ra:2013kx,younsikraNatComm,PhysRevLett.107.080504,Y.:2011aa,Nagata04052007}.  The simplest example is provided by $M=N=2$.  For twin-Fock states, the phase-dependent summand (ii) in (\ref{generalexpressionprobability}) is absent,  and we find 
\eq 
\mathcal{P}^{\ket{2,2}}_{(0,4)} &= & \frac{1}{16} \left(1 + 4  \left\{ |\braket{\phi}{\tilde \phi}|^2 \right\} +  \left\{ |\braket{\phi}{\tilde \phi}|^4 \right\}  \right) , \\ 
\mathcal{P}^{\ket{2,2}}_{(1,3)} &= & \frac 1 4 \left(1 -  \left\{ |\braket{\phi}{\tilde \phi}|^4 \right\} \right) , \\ 
\mathcal{P}^{\ket{2,2}}_{(2,2)} &= & \frac 1 {8} \left(3 - 4  \left\{ |\braket{\phi}{\tilde \phi}|^2 \right\} + 3  \left\{ |\braket{\phi}{\tilde \phi}|^4 \right\} \right)  ,
\en
 which match the results obtained via the orthonormalization of single-particle wave-functions \cite{Ra:2013kx,younsikraNatComm}. The dependence of event probabilities on different powers of $|\braket{\phi}{\tilde \phi}|^2$ stems from different bosonic exchange processes [Fig.~\ref{Feynmandiagrams.pdf}(e)]. The absence of a second-order term in $\mathcal{P}_{(1,3)}^{\ket{2,2}}$ is responsible for the narrowing of the width of the (1,3)-signal \cite{younsikraNatComm} with respect to the single-photon coherence length, the alternating signs in $\mathcal{P}_{(2,2)}^{\ket{2,2}}$ induce the non-monotonicity of the (2,2)-signal \cite{Ra:2013kx}. 

\begin{figure}[th] \center
\includegraphics[width=.9\linewidth,angle=0]{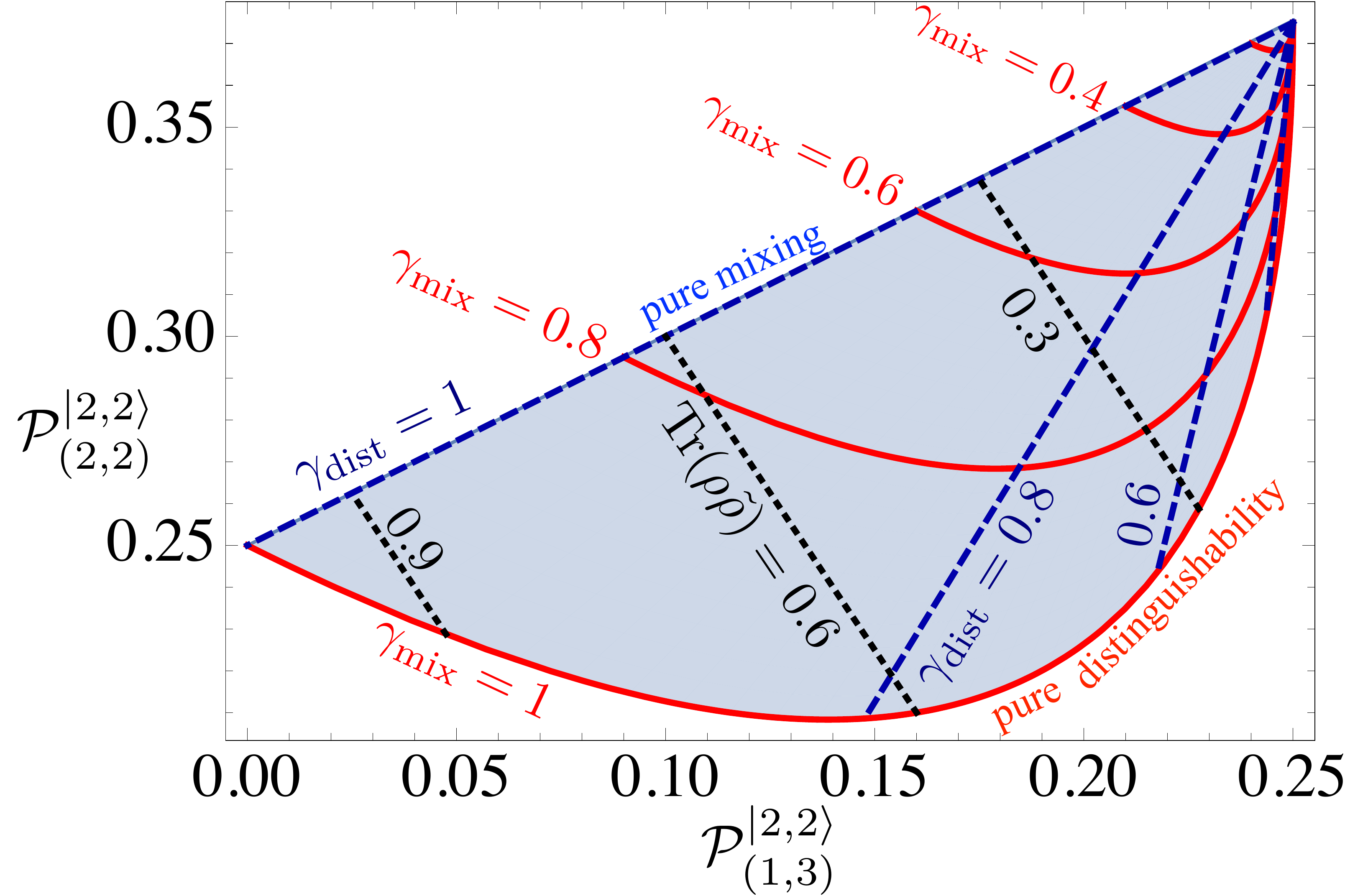}
\caption{Physical range of $(\mathcal{P}^{\ket{2,2}}_{(1,3)}, \mathcal{P}^{\ket{2,2}}_{(2,2)})$ in twin-Fock state interferometry with $\ket{2,2}$. Red solid lines show constant mixing (${\gamma_{\text{mix}}=1,0.8,0.6,0.4}$), blue dashed lines denote constant distinguishability (${\gamma_{\text{dist}}=1,0.8,0.6,0.4}$). Perfect interference takes place for $(\mathcal{P}^{\ket{2,2}}_{(1,3)}, \mathcal{P}^{\ket{2,2}}_{(2,2)})=(0,1/4)$, classical behavior is characterized by $(\mathcal{P}^{\ket{2,2}}_{(1,3)}, \mathcal{P}^{\ket{2,2}}_{(2,2)})=(1/4,3/8)$. The black dotted lines denote a constant depth of the two-particle Hong-Ou-Mandel dip, characterized by $\text{Tr}(\rho \tilde \rho)=0.9,0.6,0.3$ [Eq.~(\ref{visiHOM})]. Each dip depth allows a certain range of $\gamma_{\text{dist}}, \gamma_{\text{mix}}$, fulfilling $\mathcal{P}^{\ket{1,1}}_{(1,1)}=(1-\gamma_{\text{dist}}^2 \gamma_{\text{mix}}^2 )/2$. The experiments reported in Refs.~\cite{Ra:2013kx,younsikraNatComm} explore the distinguishability-induced quantum-to-classical transition, corresponding here to the red solid line with $\gamma_{\text{mix}}=1$. } 
  \label{diagnose4HOM} \end{figure}

Under the decoherence model above, we use (\ref{impactsp}), in addition to Eq.~(\ref{visiHOM}), we have
\eq 
\left\{ |\braket{\phi}{\tilde \phi}|^4  \right\} &=&  \gamma_{\text{dist}}^4  \gamma_{\text{mix}}^2  ,
\en
which allows us to read off $\gamma_{\text{dist}}$ and $\gamma_{\text{mix}}$ from the combined signal $(\mathcal{P}^{\ket{2,2}}_{(1,3)}, \mathcal{P}^{\ket{2,2}}_{(2,2)})$ (note that $\mathcal{P}^{\ket{2,2}}_{(0,4)}$ is fixed by the latter two), as illustrated in Fig.~\ref{diagnose4HOM}. 

As a result, the quantum-to-classical transitions induced by mixing and by distinguishability differ strongly: Pure mixing (in general: $p_j=\tilde p_j$ and $\ket{\psi_j}= \ket{\tilde \psi_j}$; here, in our model: $\gamma_{\text{dist}}=1$) implies that ASPPs of higher powers $\{ |\braket{\phi}{\tilde \phi}|^{2m} \}$ all take  the same value, saturating the upper bound of Eq.~(\ref{jensen}). Mixing therefore always induces a linear interpolation between quantum and classical probabilities, as evident from Eq.~(\ref{generalexpressionprobability}) [straight blue dashed line denoted by $\gamma_{\text{dist}}=1$ in Fig.~\ref{diagnose4HOM}]. In contrast, pure distinguishability (i.e.~the particles are described by pure states $\ket{\phi}$ and $\ket{\tilde \phi}$, the lower bound of Eq.~(\ref{jensen}) is saturated; here, $\gamma_{\text{mix}}=1$), leads, in general, to more intricate, non-monotonic transitions \cite{Ra:2013kx} [curved red solid line for $\gamma_{\text{mix}}=1$ in Fig.~\ref{diagnose4HOM}]. This  qualitative difference between these two decoherence mechanisms reinforces the role of  non-monotonicity as a  witness of a distinguishability-induced quantum-to-classical transition \cite{1367-2630-16-1-013006,ratichycomment}. 

\subsection{Double-Fock superposition $\ket{2:1}$}
A clear and unambiguous differentiation of distinguishability, mixing and dephasing is possible using the double-Fock superposition $\ket{2:1}$.  Such  a state can be generated experimentally by annihilating a single photon in a twin-Fock state $\ket{2,2}$, where the photon is extracted from either mode with the same probability \cite{YSRaPreparation}, a technique that was experimentally demonstrated for the $\ket{3:1}$ state \cite{PhysRevLett.107.163602}.    Using (\ref{generalexpressionprobability}), the two pertinent probabilities become
\eq 
\mathcal{P}^{\ket{2:1}}_{(0,3)} &=& \frac{1}{8} \left( 1  \mp \{ |\braket{\phi}{\tilde \phi}|^2  \braket{\phi}{\tilde \phi} \}\sin (\eta )  \right) \label{p2103}  \\
&& + \frac 1 4 \left( \{ |\braket{\phi}{\tilde \phi}|^2 \}   \mp  \{ \braket{\phi}{\tilde \phi} \} \sin (\eta ) \right) , \nonumber  \\
\mathcal{P}^{\ket{2:1}}_{(1,2)}&=&  \frac{3}{8} \left( 1 \pm  \{ |\braket{\phi}{\tilde \phi}|^2 \braket{\phi}{\tilde \phi} \}  \sin (\eta ) \right) \label{p2112}  \\
&& - \frac 1 4 \left(\{ |\braket{\phi}{\tilde \phi}|^2 \}  \pm  \{ \braket{\phi}{\tilde \phi} \} \sin (\eta )  \right)   \nonumber , 
\en
where the upper signs refer to the event $(s_1,s_2)$ and the lower ones to $(s_2,s_1)$. The dependence on three different ASPPs can be understood from Fig.~\ref{Feynmandiagrams.pdf}(d):  The phase-independent classical contribution is modified by a phase-independent bosonic exchange contribution, weighted by $\{ |\braket{\phi}{\tilde \phi}|^2 \} $, and two differently weighted phase-dependent terms, corresponding to exchange of one or three particles.

The three independent observables that characterize interference are the visibility of $(3,0)$-signals $\mathcal{V}^{\ket{2:1}}_{(3,0)}$, the visibility of $(2,1)$-signals $\mathcal{V}^{\ket{2:1}}_{(2,1)}$ and the total probability to find all particles in one mode, $\mathcal{P}^{\ket{2:1}}_{30+03}= \mathcal{P}^{\ket{2:1}}_{(3,0)}+ \mathcal{P}^{\ket{2:1}}_{(0,3)}$ (the total probability to find the particles in the (2,1) or (1,2)-channel is the complement $1-\mathcal{P}^{\ket{2:1}}_{30+03}$). An additional binary degree of freedom is the phase relation between the $(1,2)$ and the $(3,0)$-signal, which is formally accounted for as follows: The visibility $\mathcal{V}^{\ket{2:1}}_{(2,1)}$ is set to its negative value when the $(1,2)$ and $(0,3)$ signals are out of phase (we excluded complex scalar products $\braket{\phi}{\tilde \phi}$, such that we never encounter phase-shifts other than 0 and $\pi$). 
We combine Eqs.~(\ref{visibility10a}) and (\ref{visiHOM}) with Eq.~(\ref{impactsp2}), 
\eq 
\left\{ |\braket{\phi}{\tilde \phi}|^2 \braket{\phi}{\tilde \phi}  \right\} &=& \gamma_{\text{dist}}^3   \gamma_{\text{phase}}  \gamma_{\text{mix}}^2  , \label{dephase}
\en
to find the observables as a function of the decoherence model parameters. 
Inserting Eqs.~(\ref{visibility10a},\ref{visiHOM},\ref{dephase}) into the probabilities (\ref{p2103}, \ref{p2112}),  we can express the decoherence parameters $( \gamma_{\text{dist}} , \gamma_{\text{phase}} , \gamma_{\text{mix}} )$ as a function of the observables $\left(\mathcal{V}^{\ket{2:1}}_{(2,1)},\mathcal{V}^{\ket{2:1}}_{(3,0)}, \mathcal{P}^{\ket{2:1}}_{30+03} \right)$.

\begin{figure}[th] \center
\includegraphics[width=\linewidth,angle=0]{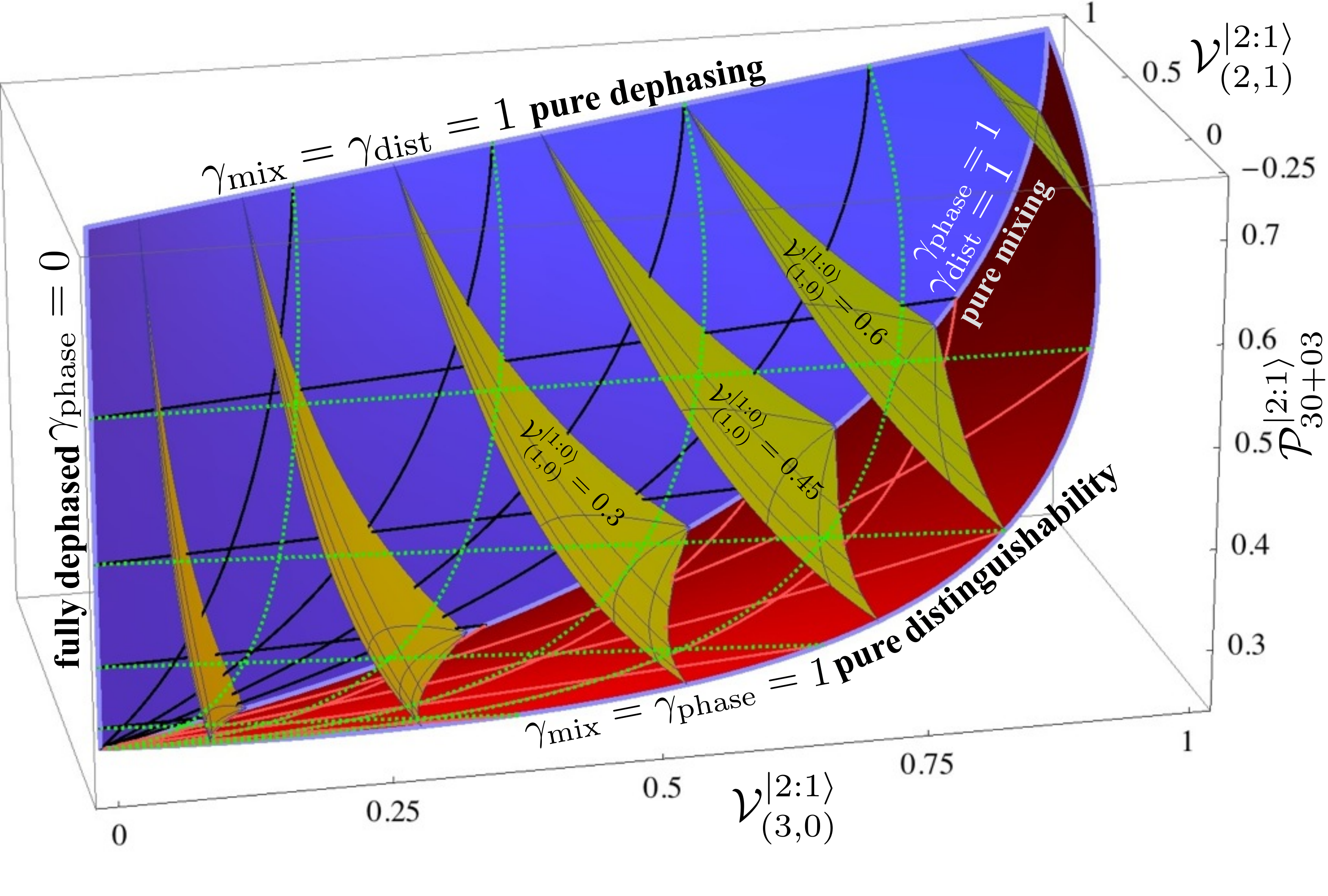}
\caption{Physical range of  $(\mathcal{V}^{\ket{2:1}}_{(2,1)},\mathcal{V}^{\ket{2:1}}_{(3,0)}, \mathcal{P}^{2:1}_{30+03})$ for the double-Fock-superposition $\ket{2:1}$. The wedge-like volume is confined by three surfaces: The blue surface in the background (with the solid black mesh) describes $\gamma_{\text{dist}}=1$, i.e.~only dephasing and mixing arise; the red surface (solid light-red mesh) corresponds to $\gamma_{\text{phase}}=1$, i.e.~only mixing and distinguishability. The light green dotted mesh in the foreground insinuates the third surface, characterized by $\gamma_{\text{mix}}=1$, i.e.~dephasing and distinguishability. The four edges (shown in light blue) correspond to pure dephasing, pure mixing, pure distinguishability, or full dephasing. The yellow surfaces are areas of constant single-particle visibility $\mathcal{V}^{\ket{1:0}}_{(1,0)}=0.05, 0.15, 0.3, 0.45, 0.6, 0.9$. Perfect interference takes place for $V^{\ket{2:1}}_{(3,0)}=V^{\ket{2:1}}_{(2,1)}=1, P^{\ket{2:1}}_{30+03}=3/4$, fully classical behavior for $V^{\ket{2:1}}_{(3,0)}=V^{\ket{2:1}}_{(2,1)}=0, P^{\ket{2:1}}_{30+03}=1/4$. 
}  \label{FullOverview.pdf}
 \end{figure}
 
The volume of physically allowed combinations of $(\mathcal{V}^{\ket{2:1}}_{(2,1)},\mathcal{V}^{\ket{2:1}}_{(3,0)}, \mathcal{P}^{\ket{2:1}}_{30+03})$ is shown in Fig.~\ref{FullOverview.pdf}. Dephasing, distinguishability and mixing lead to very different trajectories, which define the edges of the volume. When one decoherence mechanism fully destroys coherence ($\gamma_{\text{dist}}=0$ or $\gamma_{\text{phase}}=0$ or $\gamma_{\text{mix}}=0$), it is not possible to differentiate the other two: Full mixing or full distinguishability ($\gamma_{\text{dist}}=0$ or $\gamma_{\text{mix}}=0$, respectively) lead to classical behavior, independently of the value of the other decoherence parameters. When phase coherence is fully lost ($\gamma_{\text{phase}}=0$), distinguishability cannot be differentiated from mixing, as also evident from Eqs.~(\ref{visiHOM}): When the expectation values of the scalar product (\ref{visibility10a}) and the third power (\ref{dephase}) both vanish,  only the product of $\gamma_{\text{dist}}$ and $\gamma_{\text{mix}}$ can be inferred. 
If the relationship between the decoherence parameters and the observables were linear, we would observe a cube-like volume in Fig.~\ref{FullOverview.pdf}; the pathological cases explain why we instead deal with a four-sided wedge. Outside the realm of full decoherence, each choice of $( \gamma_{\text{dist}} >0 , \gamma_{\text{phase}} >0, \gamma_{\text{mix}} >0) $ leads to exactly one point $( \mathcal{V}^{\ket{2:1}}_{(2,1)}, \mathcal{V}^{\ket{2:1}}_{(3,0)}, \mathcal{P}^{\ket{2:1}}_{30+03})$, i.e.~decoherence rates can be inferred unambiguously from the observables, and differential diagnosis is possible [Fig.~\ref{Diagnosis.pdf}(b,c)]. In particular, each value of the single-particle visibility $\mathcal{V}^{\ket{1:0}}_{(1,0)}$ resulting from a single-particle interference experiment is compatible with a surface in the three-dimensional space (yellow surfaces in Fig.~\ref{FullOverview.pdf}), the exact position on that surface then clearly reveals all three decoherence parameters. Remarkably, the visibility of the $(2,1)$-signal $\mathcal{V}^{\ket{2:1}}_{(2,1)}$ can fully vanish for non-vanishing values of the decoherence parameters, due to the competition of the different phase-dependent terms of opposite sign in Eq.~(\ref{p2112}) \cite{YSRaPreparation}.  In general, full dephasing does not lead to the classical behavior of distinguishable particles: Even though both visibilities vanish for $\gamma_{\text{phase}} \rightarrow 0$,  bosonic statistics survive, favouring the $(3,0)$-channel over the $(2,1)$-channel \cite{Tichy:2012NJP}. 

\subsection{General diagnosis}
For the state $\ket{2:1}$, the three measured observables match the three physical parameters of the decoherence model presented in Section \ref{decomodel}, such that the latter can be extracted with confidence outside pathological cases. This bijective relationship, however, is not a trivial artifact of scaling to larger particle numbers: In $N00N$-state interferometry, one also measures several independent signals, but due to the unique dependence on $\{ \braket{\phi}{\tilde \phi}^N \}$ [Eq.~(\ref{N00Nstatesignal})], different decoherence processes cannot be distinguished. 

Decoherence processes that act on many particles may impact on the many-body density matrix in a complex fashion,  beyond the three-parameter model of Section \ref{decomodel}: The mixing process may affect the upper and lower arm differently and act in a more intricate way than by the addition of white noise and dephasing can occur in a non-linear fashion that impinges on different particle numbers in a different way. Moreover, non-ideal beam splitters, particle loss and imperfect detectors with finite detection efficiency and dark counts will additionally degrade the measured signals. As a general framework, a decoherence model predicts the ASPPs $\{ |\braket{\phi}{\tilde \phi}|^{2m} \braket{\phi}{\tilde \phi}^k \}$ as a function of its model parameters. 

By increasing the number of particles $N$ and $M$, we can control a larger set of observables, which allows us to keep up with the complexity of more sophisticated decoherence models and eventually infer the model parameters: The double-Fock superposition $\ket{N:M}$ yields signals that permit to infer $2M+1$ different ASPPs [Eq.~(\ref{generalexpressionprobability})]. We checked for $N=M+1$  that all $2M+1$ independent ASPPs can be inferred unambiguously from experimental observables up to $M=11$. For twin-Fock states $\ket{N,N}$, all powers $\{ |\braket{\phi}{\tilde \phi}|^{2m} \}$ for $m=1 \dots N$ can be inferred, which we checked up to $N=10$. It remains open, however, whether the relationship between experimental observables and ASPPs is always invertible.

\section{Conclusions}
Many-boson states of the form $\ket{N:M}$ provide remarkable features: Due to the dependence of event probabilities on several powers of scalar products inherent to (\ref{generalexpressionprobability}), the experimental observables are sensitive to the actual decoherence mechanism. Such double-Fock superpositions therefore provide an inexpensive way to diagnose the processes that deteriorate interferometric power. In principle, any interferometer -- be it optical, atomic or molecular -- can be diagnosed by feeding it with double-Fock-superpositions and analyzing the resulting visibilities. The alternative to differential decoherence diagnose is quantum process tomography \cite{PhysRevLett.78.390}. Since the internal state of the particle $\ket{\phi}$ typically lives in a high-dimensional Hilbert-space, such reconstruction of the full density matrix is infeasible in the current scenario.  

For large molecules \cite{MacroscopicityOpto,RevModPhys.84.157}, the current paradigm for decoherence, double-Fock superpositions are admittedly extremely challenging to generate, let alone to interfere and detect. We may alternatively gain better insight into decoherence processes with the help of other physical systems: Cold atoms in few-well-lattices provide a feasible means to test the discussed effects, since granular two-particle Hong-Ou-Mandel interference has recently been demonstrated \cite{Kaufman18072014} and cold atoms can be subject to various decoherence mechanisms in a controllable way. With photons, the three decoherence processes discussed above can be simulated by using the polarization as the distinguishing degree of freedom, and artificially inducing mixing, e.g.~in the path delay. On the other hand, the discussed methods may also help to characterize single-photon sources in a more precise way than by the usual Hong-Ou-Mandel dip \cite{Spagnolo:2013fk}, which, as we have shown in Section \ref{snf}, does not reveal the cause of imperfect interference.  As a further extension, the diagnostic power of double-Fock superpositions may also be used as a probe for other processes, for example, to quantify the non-Markovianity of an environment \cite{haikka}.

In practice, only a finite number of events can be observed, leaving the visibilities uncertain, while the mapping between model parameters and experimental observables -- with the ASPPs as intermediate step -- might be quite intricate. Such more complex scenarios can be treated via Bayesian methods, which may also allow to design optimized measurement strategies to quickly and reliably reveal the actual values of decoherence parameters \cite{PhysRevLett.99.223602,PhysRevLett.100.073601}. Using double-sided Feynman diagrams, our analysis can be  taken further to general states of the form $\ket{\Phi(\vec \alpha)}=\sum_{n=0}^{N_{\text{tot}}}  \alpha_n \ket{n,N_{\text{tot}}-n}$. On the one hand, the  $\vec \alpha=\{ \alpha_1, \dots \alpha_{N_{\text{tot}}} \}$  can be adjusted  to achieve the best sensitivity to the type of decoherence process, i.e.~the best differential diagnosis. On the other hand, given a fully diagnosed interferometer, the optimal $\vec \alpha$ that achieves the best phase-sensitivity \cite{PhysRevA.76.013804} may itself depend on the actually occurring decoherence processes.  
It remains to be studied to which extent the methods of \cite{PhysRevLett.98.223601,PhysRevA.87.053821,PhysRevLett.109.233603}, which rely on post-selecting a desired output state in order to synthesize phase-super-resolving interference signals, can be extended to the present purpose of decoherence diagnosis. 
From a more fundamental perspective, the complicated dependence of visibilities on decoherence parameters challenges any attempt to formulate a complementarity relation \cite{bertet2001,Bose:2002vf,Scully:1991uq,jakob2007cae} between particle-like and wave-like behavior as well as to  quantify macroscopic interference \cite{MacroscopicityOpto,froewisduer}, which remain great desiderata. 

\subsection*{Acknowledgements}
M.C.T. would like to thank Pinja Haikka, Steven Kolthammer, Benjamin Metcalf and Ian Walmsley for inspiring discussions and useful comments, and Christian Kraglund Andersen, Alexander Holm Kiilerich, Robert Keil, Andrew Wade and Qing Xu for very valuable feedback on the manuscript.  M.C.T and K.M. acknowledge funding by the Villum foundation and by the Danish Council for Independent Research. H.-T.L. acknowledges the financial support from the National Junior Research Fellowship (Grant No. 2012- 000642). C.G. acknowledges financial support by DAAD. Y.-H.K. acknowledges financial support by the National Research Foundation of Korea (Grant No. 2013R1A2A1A01006029).

\appendix

\begin{figure}[th] \center
\includegraphics[width=.75\linewidth,angle=0]{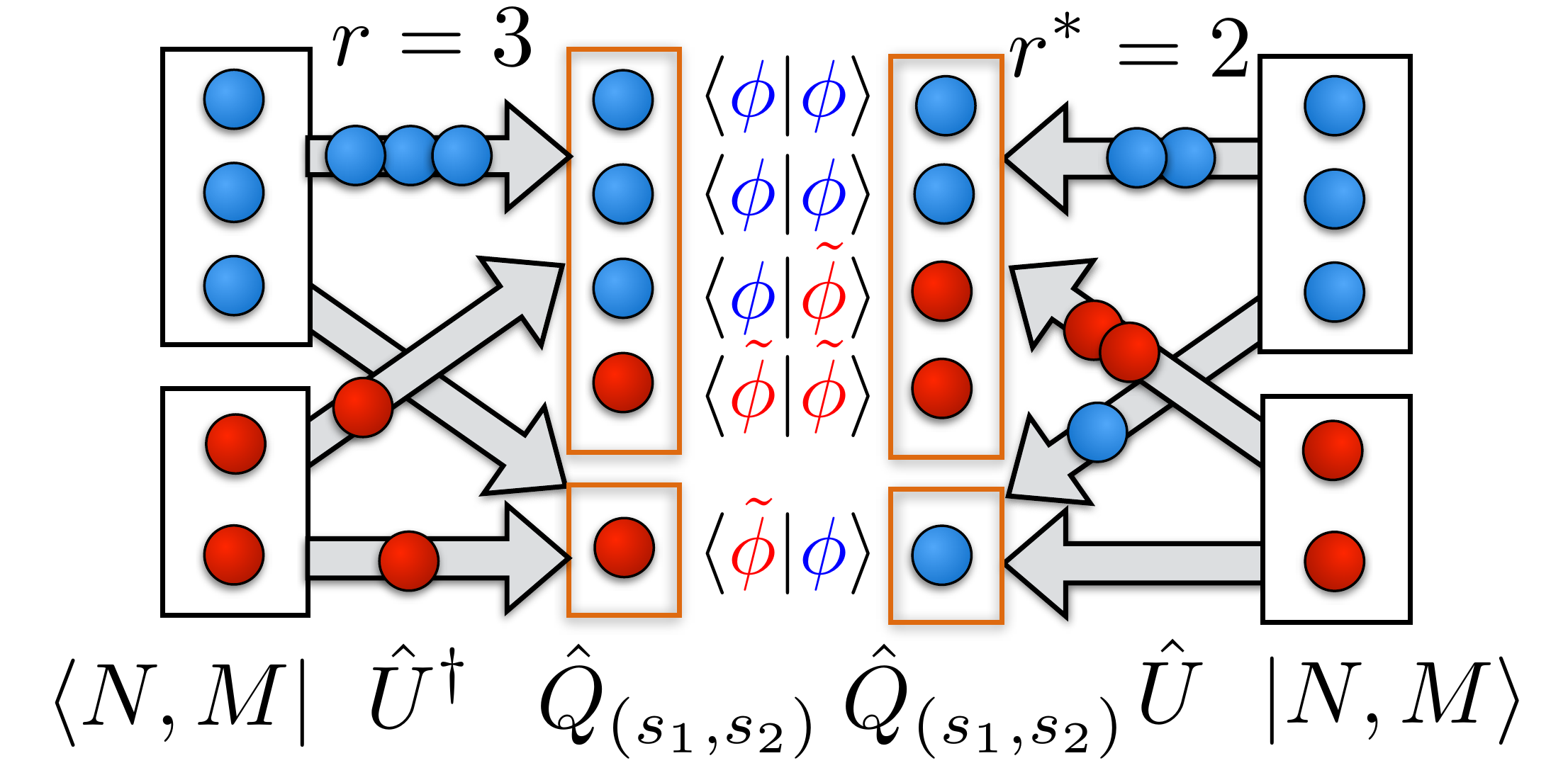}
\caption{ One  summand of Eq.~(\ref{cdef}): We set $N=3, M=2$, $s_1=4, s_2=1$, and consider the process with $r^*=2$ and $r=3$ reflected particles from the first input mode in time-forward and time-backward direction, respectively. There are $J=1$ pairs of exchanged particles, leading to a weight $|\braket{\phi}{\tilde \phi}|^2$, and one ($j=1$) exchange occurs in the first output mode. 
 }  \label{figappendix.pdf}
 \end{figure}

\begin{widetext}
\section{Computation of event probabilities} \label{eventprobapp}
The coefficient $\mathcal{C}_J$ in Eq.~(\ref{generalexpressionprobability}) is given by 
\eq 
C_J &=& \sum_{r, r^*=\text{max}(0,s_1-M) }^{\text{max}(0,s_1-M) +\text{min}(M,s_2)  }  (-1)^{r+r^*}
 \sum_{\substack{
 j=\text{max}(0,2J-s_2), \\  j \stackrel{!}{=} r-r^*\mod 2 }  }^{\text{min}(s_1,2J)} 
\mathcal{M}^{(r,s_1-r)}_{(r^*, s1-r^*)}(j)  \mathcal{M}^{(N-r,s_2-N+r)}_{(N-r^*, s_2-N+r^*)}(2J-j)  ,  \label{cdef}
\en
where
\eq
  \mathcal{M}^{p,n-p}_{q,n-q}(j) & =&  {n \choose p} {p \choose (j+p-q)/2 } {n-p \choose (j-p+q)/2 }  \label{mdef}. 
\en
\end{widetext}
The sum Eq.~(\ref{cdef}) is illustrated in Fig.~\ref{figappendix.pdf} and can be interpreted as follows: The particles are redistributed from the input to the output modes, with $r$ ($N-r$) particles from the first input mode found in the first (second) output mode.   In order to eventually measure $s_1$ and $s_2$ particles in the first and second output modes, respectively,  $s_1-r$ ($s_2-(N-r)$) particles from the second  input mode must be found in the first (second) output mode. Since only non-negative particle numbers are allowed for the four processes, $r$ is restricted to a certain range of values. 
To yield the probability Eq.~(\ref{prob1st}), we 
 remain with two sums, over $r$ and $r^*$. The relative phase acquired by such a process is $(-1)^{r+r^*}$, an additional relative phase arises for the exchange term (ii) in Eq.~(\ref{generalexpressionprobability}).  The sum over $j$ accounts for the bosonic exchange processes in the first output mode, i.e.~$j$  $\ket{\phi}$-particles are exchanged with $ \ket{ \tilde \phi}$-particles; consequently, $2J-j$ exchange processes occur in the second output mode. The overnormalization due to the multiple creation of bosons in the same mode is accounted for by $ \mathcal{M}^{p,n-p}_{q,n-q}(j)$. 
 
 Eq.~(\ref{cdef}) can alternatively be derived using a decomposition of single-particle wave-functions in an orthonormal basis \cite{tichyTutorial,Ra:2013kx,younsikraNatComm}, for which, however, the clear separations in combinatorial factors and scalar products in  Eq.~(\ref{generalexpressionprobability}) only emerges after lengthy algebraic manipulations.

\end{document}